\documentstyle[aps,prd,epsf]{revtex}

\begin{document}


\newcommand{\be}{\begin{equation}}
\newcommand{\ee}{\end{equation}}
\newcommand{\ben}{\begin{eqnarray}
\displaystyle}
\newcommand{\een}{\end{eqnarray}}

\newcommand{\la}{{\lambda}}
\newcommand{\Si}{{\Sigma}}
\newcommand{\de}{{\delta}}
\newcommand{\tde}{{\tilde \delta}}

\newcommand{\C}{{\cal C}}
\newcommand{\cP}{{\cal P}}
\newcommand{\cA}{{\cal A}}
\newcommand{\cB}{{\cal B}}
\newcommand{\cE}{{\cal E}}
\newcommand{\cH}{{\cal H}}
\newcommand{\cJ}{{\cal J}}
\newcommand{\cJn}{{\cal J}_{\infty}}
\newcommand{\cO}{{\cal O}}
\newcommand{\cQ}{{\cal Q}}
\newcommand{\cR}{{\cal R}}
\newcommand{\cS}{{\cal S}}
\newcommand{\cT}{{\cal T}}
\newcommand{\cU}{{\cal U}}
\newcommand{\cY}{{\cal Y}}
\newcommand{\cZ}{{\cal Z}}
\newcommand{\cL}{{\cal L}}
\newcommand{\M}{{\cal M}}

\newcommand{\p}{\partial}
\newcommand{\na}{\nabla}
\newcommand{\ints}{\int_{\Sigma} d\Sigma}
\newcommand{\LieN}{{\cal L}_{N^{i}}}
\newcommand{\Lief}{{\cal L}_{\phi^{i}}}
\newcommand{\Liet}{{\cal L}_{t^{i}}}
\newcommand{\LieM}{{\cal L}_{M^{\mu}}}
\newcommand{\Lie}{{\cal L}}

\newcommand{\tiA}{{\tilde A}}
\newcommand{\tiB}{{\tilde B}}
\newcommand{\tim}{{\tilde \mu}}
\newcommand{\tir}{{\tilde r}}
\newcommand{\trp}{{\tilde r_{+}}}
\newcommand{\hr}{{\hat r}}
\newcommand{\rv}{{r_{v}}}
\newcommand{\dr}{{d \over d \hr}}

\newcommand{\hhf}{{\hat \phi}}
\newcommand{\hhM}{{\hat M}}
\newcommand{\hhZ}{{\hat Z}}
\newcommand{\hhS}{{\hat \Sigma}}
\newcommand{\hhD}{{\hat \Delta}}
\newcommand{\hhm}{{\hat \mu}}
\newcommand{\hro}{{\hat \rho}}
\newcommand{\hhz}{{\hat z}}
\newcommand{\tF}{{\tilde F}}
\newcommand{\hT}{\hat T}
\newcommand{\tT}{\tilde T}
\newcommand{\hC}{\hat C}

\newcommand{\ep}{\epsilon}
\newcommand{\bep}{\bar \epsilon}
\newcommand{\Ga}{\Gamma}
\newcommand{\ga}{\gamma}
\newcommand{\hth}{\hat \theta}

\title{Extrema of Mass, First Law of Black Hole
Mechanics and Staticity Theorem in
Einstein-Max\-well-axi\-on-di\-la\-ton Gravity}

\author
{Marek Rogatko}
\address{
Technical University of Lublin \protect \\
20-618 Lublin, Nadbystrzycka 40, Poland \protect \\
rogat@tytan.umcs.lublin.pl \protect \\
rogat@akropolis.pol.lublin.pl}

\date{\today}

\maketitle

\begin{abstract}
Using the ADM formulation of the Einstein-Maxwell axion-dilaton gravity
we derived the formulas for the variation of mass and other asymptotic
conserved quantities in the theory under consideration. Generalizing
this kind of reasoning to the initial data for the manifold with
an interior boundary we got the generalized first law of black hole
mechanics. \par
We consider an asymptotically flat solution to the
Einstein-Maxwell axion-di\-la\-ton gravity 
describing a black hole with a Killing vector field timelike at
infinity, the horizon of which comprises a bifurcate Killing horizon with
a bifurcate surface. Supposing that the Killing vector field is
asymptotically orthogonal to the static hypersurface with boundary S
and a compact interior, we find that the solution is static
in the {\it exterior world}, when the
timelike vector field is normal to the horizon and has vanishing {\it
electric} and {\it axion-electric} fields on static slices.
\end{abstract}

\smallskip

\baselineskip=18pt
\par
\section{Introduction}
During the last years the discovery of new black hole 
solutions in theories with nonlinear matter fields
(see only a selected list of references \cite{ne})
prompted us to study topics related to the stationary problem of nonrotating
black holes as well as the subject of stationarity of these objects.
Nowadays, the problems of black hole physics of the late 1960s and
1970s are
reassessed, taking into consideration nonlinear matter models or
general sigma models.
\par
From the historical point of view
the idea of a staticity theorem was proposed by Lichnerowicz \cite{li}
for the simple case in which there was no black hole. He proved 
that case for a
stationary perfect fluid that was everywhere locally static in the sense 
that its flow vector was aligned with the Killing vector. The Killing vector
itself would have the staticity property of being hypersurface orthogonal.
\par
The next extension of the research was attributed to Hawking \cite{ha}.
His  proof of staticity applied to the vacuum case. He
considered black holes that were 
nonrotating in the sense that the null generator
of the horizon was aligned with the Killing vector.
Carter considered
an extension of this problem to the case of electromagnetic fields and obtained the desired
result subject to a
certain inequality
between the norm of the Killing field and the electric potential
\cite{ca0,ca}.
\par
By means of the Arnowitt-Deser-Misner (ADM) formalism, 
Sudarsky and Wald \cite{su} considered an asymptotically flat solution
to Einstein-Yang-Mills (EYM) equations with a Killing vector field which was
timelike at infinity. By means of the notion of an asymptotically flat
maximal slice with compact interior, they established that the solution
is static when it had a vanishing Yang-Mills electric field on the static
hypersurfaces. If an asymptotically flat solution possesses a black
hole, then it is static when it has a vanishing electric field on the
static hypersurface. They also presented a new derivation of the mass formula and
proved that every stationary solution is an extremum of the ADM mass
at fixed Yang-Mills electric charge. On the other hand,
every stationary black hole solution is an extremum of the ADM mass
at fixed electric charge, canonical angular momentum, and horizon area.
One should also mention the work of Sudarsky and Wald
\cite{su1}, in which they derived new integral mass
formulas for stationary black holes in EYM theory.
Using the notion of maximal hypersurfaces \cite{ch} and
combinig the mass formulae,
they obtained the proof that nonrotating Einstein-Maxwell (EM)
black holes must be static and have a vanishing magnetic field 
on the static slices.
\par
The {\it strong rigidity theorem} \cite{hab}, derived by Hawking,
emphasizes that the event horizon of a stationary
black hole had to be a Killing horizon;
i.e., there had to exist a Killing field $\chi_{\mu}$
in the spacetime which was normal to the horizon. If this 
field did not coincide
with the stationary Killing field $t_{\mu}$,
then it was shown that the spacetime had to be axisymmetric as well as 
stationary.
It follows that the black hole will be rotating;
i.e., its angular velocity of the horizon $\Omega$ will be nonzero
($\Omega$ is defined by the relation $\chi_\mu = t_\mu + \Omega \phi_\mu$,
where $\phi_\mu$ is an axial Killing vector field) and the 
Killing vector field $k_\mu$
will be spacelike in the vicinity of the horizon. 
The black hole will be enclosed by an
ergoregion.
On the other hand, if $t_\mu$ coincides with $\chi_\mu$ (so that the
black hole is
nonrotating) and $t_\mu$ is globally timelike outside the black hole,
then one can show that the spacetime is static.
The standard black hole uniqueness theorem leaves an open question 
of the problem
of the potential existence of additional stationary black hole
solutions of EM equations 
with a bifurcate horizon which are neither static nor axisymmetric.
The situation was recuperated by Wald \cite{www}. 
He showed that any
nonrotating black hole in EM theory, the ergoregion of which
was disjoint 
from the horizon had, to be static, even if the $t_{\mu}$ was not initially 
presupposed to be globally timelike outside the black hole.
\par
Chru\'sciel \cite{rig} reconsidered the problem of
{\it the strong rigidity theorem} and gave the corrected version of the
theorem
in which he excluded the previous
assumption about
maximal analytic extensions which were not unique.

\par
\medskip
The uniqueness theorems for black holes are closely related to the
problem of staticity. However, the uniqueness theorems are based on 
stronger assumptions than
the {\it strong rigidity theorem}. Namely, 
in the nonrotating case one requires staticity whereas in the  rotating case
the uniqueness theorem is established for circular spacetimes. 
The foundations of the uniqueness theorems were laid by
Israel \cite{is,is1} who established the 
uniqueness of the Schwarzshild metric 
and its Reissner-Nordstr\"om generalization
as static asymptotically flat solutions of the Einstein and EM
vacuum field equations. Then, 
M\"uller zum Hagen {\it et al.} \cite{mi} 
in their works were able to 
weaken Israel's assumptions concerning the topology and regularity
of the two-surface $V = - t^\mu t_\mu = const$.
Robinson
\cite{ro} generalized the theorem of
Israel concerning the uniqueness of the Schwarzschild black hole \cite{is}.
Finally, Bunting and Masood-ul-Alam \cite{bu} excluded 
multiple black hole solutions,
using the 
conformal transformation and the
positive mass theorem \cite{po}. Lately, a generalization of the results to 
electrovacuum spacetimes was achieved \cite{la}.
\par
The uniqueness results for rotating configurations, i.e., for
stationary, axisymmetric black hole spacetimes were obtained by 
Carter \cite{cr},
completed by Hawking and Ellis \cite{hab} and the next works of Carter
\cite{ca0,ca} and Robinson \cite{rr}. They were related to the vacuum case.
Robinson also gained \cite{rr1} a complicated indentity which enabled him to
expand Carter's results to electrovac spacetimes.
\par
A quite different approach to the problem under consideration was presented by
Bunting \cite{br} and Mazur \cite{mr}.
Bunting's approach was based on applying a general 
class of harmonic mappings between
Riemannian manifolds while Mazur's was based on the observation that the Ernst
equations describe nonlinear $\sigma$ model on symmetric space.
A review of these new methods presented by Bunting and Mazur was 
given in Ref.\cite{ccr}.
\par
A recent review which covers in detail various 
aspects of the uniqueness theorems for nonrotating and rotating
black holes was provided by 
Heusler \cite{book}.
\par
Heusler and Straumann in Ref.\cite{str1} 
considered the stationary EYM and Einstein dilaton
theories. They showed
that the mass variation formula involves only global quantities and surface
terms; their results hold for arbitrary gauge groups and any structure
of the Higgs field 
multiplets. In Ref.\cite{str} the same authors studied the staticity 
conjecture and 
circularity conditions for rotating black holes in EYM
theories. It turned out that contrary to the Abelian case staticity conjecture
might not hold for non-Abelian gauge fields like
the circularity theorem for these fields.
Recently, it has been shown \cite{nab} that in the non-Abelian case
stationary balck hole spacetimes with vanishing angular momentum need
not to be static unless they have vanishing electric Yang-Mills charge.
Heusler  \cite{si} demonstrated that any selfcoupled, 
stationary scalar mapping (${\sigma}$ model)
from a domain of strictly outer communication, with nonrotating
horizon, has to be static.
He also proved no-hair conjecture for this model.
\par
The mathematical rigor of the uniqueness theorems and related topics were
subject to the review articles by Chru\'sciel \cite{ch1,ch2}.

\medskip
Presently it seems that the most promising candidates 
for a theory of quantum gravity
are strings theories. Their implications on the theory of gravity 
and on black holes are widely elucidated. 
In what follows, we will concentrate on the
Einstein-Maxwell axion-dilaton  (EMAD) model, which is relevant to the bosonic
sector of heterotic string theory.
\par
Much work has been 
devoted to the so-called axion-dilaton gravity which is the truncation of
$N = 4$, $d = 4$ supergravity with only one vector field Refs.
\cite{ka1} - \cite{ke}. 
\par
Gal'tsov studied \cite{ga4}
axion-dilaton gravity interacting with $p$-$U(1)$ vectors in 
four-dimensional
spacetime admitting a non-null Killing vector.
A new set of supersymmetric stationary solutions of pure $N = 4, d = 4$
supergravity, generalizing the Israel-Wilson-Perj\'es solution of
EM theory was presented by Bergshoeff, Kallosh and Ortin \cite{bka}.
The authors argued that one vector field is insufficient to generate
all the interesting metrics.
\par
Rogatko \cite{rog} derived the general Smarr formula and general variation
formula for stationary axisymmetric black holes in 
EMAD gravity.
Heusler \cite{heno} pointed out that various
self-gravitating field theories with massless scalars and vector fields
reduce to the $\sigma$ models, effectively coupled to three-dimensional
gravity. Using the coset structure to construct conserved currents and
closed two-forms, integrating the latter over a spacelike hypersurface,
he provides a generalized Smarr formula for stationary black holes
with nonrotating Killing horizon, for both EM and
EMAD systems.
\par
Using the canonical formalism for the theories with the matter content
arising in string theory, Larsen and Wilczek in Ref.\cite{lar} 
studied the problem
of classical hair in string theory. They derived an effective theory
for the hair in terms of the horizon variables. It has turned out that
the solution of the constraints expresses these variables in terms of
hair seeing by an observer at infinity.
\par
Gibbons {\it et al.} \cite{l1} derived the first law of black hole
termodynamics by means of the variation of moduli fields. They have
shown that the ADM mass is extremized at fixed area, angular momentum,
and electric and magnetic charges when the moduli fields took the fixed
values depending on electric and magnetic charges. It follows from
their research that at least mass of any black hole with fixed
conserved electric and magnetic charges is given by the mass of the
double-extreme black hole with these charges.
\par
One should also mention the work of Creighton and Mann \cite{man}
in which they considered gravity coupled to various types of Abelian
and non-Abelian gauge fields with three and four-form field strenghts.
Using the quasilocal formalism \cite{york} they found the entropy for
stationary black holes and derived the first law of black hole
thermodynamics for black holes with the gauge fields under consideration.
\par
The problem of staticity in the theory under consideration
was studied by Rogatko in Ref.\cite{rog3}.
Using the modified Carter arguments it was proved that the condition of
vanishing the asymptotic value of the quantity assembled by means of 
$SL(2,R)$ duals to the $U(1)$ gauge fields and satisfying the inequality
for the sum of potentials 
enabled to satisfy the staticity conditions for fields
and metric. This inequality should hold everywhere in the domain of outer
communication. However, this assumption has no physical justification.
The big challange will be to eliminate the aforementioned assumption
from the considerations.
\par
Our paper is organized as follows. In Sec.II, we present the
canonical formalism for the EMAD gravity.
Then, in Sec.III we introduce the exact form of the canonical
energy and angular momentum in the theory under consideration. We will
show that every stationary solution in EMAD
gravity is an extremum of the ADM mass at fixed {\it dilaton-electric}
charge. We also derived the first law of
the black holes dynamics and drew a conclusion that any stationary
black hole with a bifurcate Killing horizon is an extremum of the ADM
mass at fixed {\it dilaton-electric},
canonical angular momentum, and horizon area. In Sec.IV, we deal with
the staticity problem for nonrotating black holes, finding the
conditions on which the solutions are static.
\par

\vspace{0.2cm}
\noindent
In our paper Greek indices will range from 0 to 3 and denote tensors on
four-dimensional manifold. On the other hand Latin indices range from 1 to
3 and denote tensors on a spatial hypersurface $\Si$. $g_{\alpha
\beta}$ will be the metric of the spacetime, while $h_{ab}$ will live
on a spatial hypersurface $\Si$. The corresponding covariant
derivatives are denoted by $\na_{\alpha}$ and $\na_{a}$.

\section{Canonical Formalism for Einstein-Maxwell-axion-dilaton
Gravity}
The heterotic strings provides an interesting generalization of the
EM theory in the so-called low energy limit. The bosonic
sector of the dimensionally reduced effective action for the string theory 
is obtained when we consider six of ten dimensions which have been
compactified \cite{sen,g,bka}. A simplified model of this kind is an
EMAD coupled system. It contains 
a metric $g_{\mu \nu}$, $U(1)$ vector fields
$A_{\mu}$, a dilaton $\phi$,
and a three index antisymmetric tensor field $H_{\alpha \beta \ga}$. 
In our work we will consider the action of the form \cite{stw}
\be
I = \int d^4 x \sqrt{-g} \left [ R - 2(\na \phi)^{2} - {1 \over 3}
e^{-4\phi} H_{\alpha \beta \ga} H^{\alpha \beta \ga}
- e^{-2\phi}F_{\alpha \beta} F^{\alpha \beta} \right ],
\label{a1}
\ee
where $F_{\mu \nu} = 2\na_{[\mu} A_{\nu]}$ and 
$H_{\alpha \beta \ga}$ stands for the three-index antisymmetric
tensor field defined by
\be
H_{\alpha \beta \ga} = \na_{\alpha} B_{\beta \ga} - A_{\alpha}
F_{\beta \ga} + cyclic.
\ee
The equations of motion corresponding to the action (\ref{a1})
derived from the variational principle are given by
\be
\na_{\mu}\left ( e^{-2 \phi} F^{\mu \alpha} \right )
+ {1 \over 2} e^{-4\phi} H^{\alpha \beta \ga} F_{\beta \ga} = 0,
\ee

\be
\na_{\mu} \left ( e^{-4\phi} H^{\mu \alpha \beta} \right ) = 0,
\label{a2}
\ee
\be
\na_{\mu} \na^{\mu} \phi + {1 \over 3}e^{-4 \phi}H_{\alpha \beta \ga}
H^{\alpha \beta \ga} + {1 \over 2}e^{-2\phi}F_{\alpha \beta}F^{\alpha \beta}
= 0.
\label{a3}
\ee

\vspace{0.4cm}
In this section, we consider the canonical formalism for the theory
under consideration.
The ADM formalism considered four-geometry to consist of a 
foliation of three-geometries. Its main idea is that the geometry of the
manifold is described in terms of the intrinsic metric and the
extrinsic curvature of a three-dimensional hypersurface, along with the lapse
function and the shift vector. The lapse and shift relate the intrinsic
coordinate on one hypersurface to the intrinsic coordinates on a nearby
hypersurface. Spacetime is sliced into spacelike hypersurfaces with
each hypersurface labeled by a global time parameter (see, e.g.,
\cite{adm,wa}).
\par
Thus, the canonical formalism divide the metric into spatial
and temporal parts, namely
\be
ds^{2} = - (N dt)^{2} + h_{ab} (dx^{a} + N^{a}dt)(dx^{b} + N^{b}dt),
\label{a4}
\ee
where general covariance implies the great arbitrarines in the choice
of lapse and shift functions $N^{\mu}(N, N^{a})$.
\par
In the canonical formulation of the EMAD theory
the point in the phase space corresponds
to the specification of the fields
$(h_{ab}, \pi^{ab}, A_{a}, E^{a}, B_{ij}, E^{ij}, \phi, E)$ 
on a three-dimensional
$\Sigma$ manifold. Here $h_{ab}$ is a Riemannian metric on $\Sigma$,
$A_{a}$ is the $U(1)$ gauge field on 
the three-dimensional manifold, $B_{ij}$ is a
Kalb-Ramond antisymmetric tensor field and $\phi$ is the dilaton field
on $\Si$.
The field momenta can be found by varying the Lagrangian
with respect to $\na_{0}h_{ab}$, $\na_{0}\phi$, $\na_{0}A_{a}$
and $\na_{0}B_{ij}$, where $\na_{0}$ denotes the derivative with
respect to t-coordinate. 
\par
Performing the variations, one has that
the momentum $\pi^{ab}$ canonically conjugate to a Riemannian metric can be 
expressed by means of the extrinsic curvature $K_{ab}$ of $\Sigma $ 
hypersurface
\be
\pi^{ab} = \sqrt{h} \left ( K^{ab} - h^{ab}K \right ).
\label{a5}
\ee
The momenta canonically conjugate to the Kalb-Ramond tensor
fields $B_{ij}$ and to the dilaton $\phi$
fields are given, respectively, as
\be
\pi_{ij}^{(H)} = 2 e^{-4\phi} E_{ij},
\label{a6}
\ee
\be
\pi^{(\phi)} = 4E,
\label{a7}
\ee
where
\ben \label{a8}
E_{ij} = \sqrt{h}H_{\mu ij} n^{\mu},\\
E = \sqrt{h}\na_{\mu} \phi n^{\mu}.
\een
$n^{\mu}$ is the unit normal timelike vector to the hypersurface $\Si$
in the spacetime. 
In what follows, for the brevity of considerations, the quantity
$E_{ij}$ will be called as {\it axion-electric} field and $E$ as
{\it dilaton-electric field}.
\par
The momentum $\pi_{i}^{(A)}$ canonically conjugate to the $U(1)$
gauge fields $A_{a}$ is equal to
\be
\pi_{i}^{(A)} = 4 e^{-2\phi} E_{i} + 4 e^{-4\phi}E_{ij} A^{j},
\label{a9}
\ee
where the {\it electric} field $E_{i}$ has the form
\be
E_{i} = \sqrt{h}F_{\mu i} n^{\mu}.
\label{a10}
\ee
The Hamiltonian, defined by the Legendre transform, yields
\ben \label{a11}
\cH &=& \pi^{ij} \na_{0} h_{ij} + \pi^{(\phi)} \na_{0}\phi +
\pi^{i(A)} \na_{0} A_{i} 
+ \pi^{ij(H)} \na_{0}B_{ij}
- \cL_{EMAD} \\ \nonumber
&=& N^{\mu} C_{\mu} + A_{0} \tiA + B_{j0} \tiB^{j} + \cH_{div}.
\een
where $\cH_{div}$ is the total derivative and has the form
\be
\cH_{div} = \na_{i}(\pi^{i(A)} A_{0}) - 4\na_{i}( e^{-4\phi} B_{j0})
+ 2 \na_{a} \left ( {N_{b} \pi^{ab} \over \sqrt{h}} \right ).
\label{a12}
\ee
The gauge field $A_{0}$ and $B_{j0}$ have no associated kinetic terms,
so one can regard them as Lagrange multipliers. They correspond
to the generalized {\it Gauss law}: namely
\ben \label{a13}
0 &=& \tiA = 2 e^{-4\phi} F_{ij} E^{ij} - \na_{i} \pi^{i(A)}, \\
0 &=& \tiB^{j} = 4 \na_{i} \left ( e^{-4\phi} E^{ij} \right ).
\een

\vspace{0.3cm}
As in Refs.\cite{su,re}, 
in our considerations we will take into account the
asymptotically flat initial data (sometimes called 
{\it the regular initial data});i.e.,
we assume that there is a region $C \subset \Si$ diffeomorphic to
$R^3 - B$ with $B$ compact and $C$ such that
the following
conditions are satisfied at infinity
\ben \label{a14}
h_{ab} \approx \de_{ab} + \cO \left ({1 \over r} \right ),\\ 
\na_{c} h_{ab} \approx \cO \left ({1 \over r^2} \right ), \\
\pi^{ab} \approx \cO \left ({1 \over r^2} \right ),\\
E \approx \cO \left ({1 \over r^2} \right ), \\
\phi \approx \cO \left ({1 \over r} \right ), \\
E_{a} \approx \cO \left ({1 \over r^2} \right ), \\
A_{b} \approx \cO \left ({1 \over r} \right ), \\
E_{ij} \approx \cO \left ({1 \over r^3} \right ),\\
B_{ij} \approx \cO \left ({1 \over r^3} \right ),
\een
where $\de_{ab}$ is a spatial metric on the hypersurface for which $t =
const$. We also assume \cite{re} the following conditions for the lapse
and shift functions at infinity:
\ben \label{sh}
N \approx 1 + \cO \left ({1 \over r} \right ), \\
N^{a} \approx \cO \left ({1 \over r} \right ).
\een
The other requirements will be that $k-th$ order derivatives of
the above quantities fall off $k$ power of $r$ faster than described by
the above equations.
\par
The EMAD gravity is a theory with constraints.
On a hypersurface $\Si$, initial data are restricted to the case
that at each point $x \in \Si$, one has
\ben \label{a15}
0 &=& C_{0} = \sqrt{h} \left [ 
- R^{(3)} + \left ( \pi_{ab} \pi^{ab} - {1 \over 2} \pi^{2} \right )
\left ( {1 \over h} \right ) \right ] + \\ \nonumber
&+& {1 \over \sqrt{h}} \left (
2 E^2 + 2 e^{-2 \phi}E_{i}E^{i} + e^{-4\phi}E_{ij}E^{ij} +
8 e^{-4\phi}E_{i}E^{ij}A_{j} \right ) + \\ \nonumber
&+& 2 \sqrt{h} \na_{a} \phi \na^{a} \phi +
\sqrt{h} e^{-2\phi} F_{ab}F^{ab} + {1 \over 3} e^{-4\phi} H_{ijk}
H^{ijk}, 
\een
\be
0 = C_{a} = - 2 \sqrt{h}\na_{b} \left ( {\pi_{a}{}{}^{b} \over
\sqrt{h}} \right ) + 4 E \na_{a} \phi +
2 e^{-4 \phi} H_{aij} E^{ij} + 4 e^{-2\phi} E^{i}F_{ai} + 8 E^{-4\phi}
F_{ai}E^{ij}A_{j},
\label{a16}
\ee
\be
0 = \tiA = 2 e^{-4\phi} F_{ij} E^{ij} - \na_{i} \pi^{i(A)},
\label{a}
\ee
\be
0 = \tiB^{j} = 4 \na_{i} \left ( e^{-4\phi} E^{ij} \right ).
\label{b}
\ee
For the completeness of the constraints equations we write here
one more time the generalized {\it Gauss law}, Eqs.(\ref{a}) and (\ref{b}).
The equations of motion, for the system under consideration, can be
formally derived from a Hamiltonian of the form
\be
H_{V} = \ints \left ( N^{\mu} C_{\mu} + N^{\mu} A_{\mu} \tiA
+ N^{\mu} B_{j \mu} \tiB^{j} \right ).
\label{a17}
\ee
Here we have used the form of the Hamiltonian proposed in
\cite{su,www} in the case of EYM system and EM gravity.
A general variation of the initial data
$(\de h_{ab}, \de \pi^{ab}, \de \phi, \de E, \de A_{a}, \de E^{a},
\de B_{ij}, \de E^{ij})$ of a compact support, will cause a variation
in Hamiltonian.
After performing integration, we reach to the formula
\be
\de H_{V} = \ints \left ( \cP^{ab} \de h_{ab} +
\cQ_{ab} \de \pi^{ab} + \cR^{a} \de A_{a} + \cS_{a} \de E^{a} +
\cT^{ij} \de B_{ij} + \cU_{ij} \de E^{ij} + \cY \de \phi + \cZ \de E \right ),
\label{ha}
\ee
where $\cP^{ab}, \cQ_{ab}, \cR^{a}, \cS_{a}, \cT^{ab}, \cU_{ab}, \cY,
\cZ$ are given in the forms
as follows
\be
\cP^{ab} = \sqrt{h} N a^{ab} + \sqrt{h} \left ( 
h^{ab} \na^{i} \na_{i} N - \na^{a} \na^{b} N \right ) -
\LieN \pi^{ab},
\label{ap}
\ee
\be
\cQ_{ab} = \left ( 2 \pi_{ab} - \pi h_{ab} \right ) {N \over \sqrt{h}} +
\LieN h_{ab},
\label{aq}
\ee
\ben
\cR^{a} = &-& 4 \sqrt{h} \na_{i} \left ( e^{-2\phi}N F^{ia} \right ) -
2 N \sqrt{h} e^{-4\phi}  H^{aij} F_{ij} +
4 \na_{i} \left ( \sqrt{h} N e^{-4\phi} H^{mia} A_{m} \right ) +
8 {N \over \sqrt{h}} e^{-4\phi} E_{i} E^{ia} + \\ \nonumber
&-& 4 \LieN (e^{-2 \phi}E^{a}) - 4 \LieN (e^{-4\phi}E^{ai}A_{i}) -
8 (\LieN A_{j}) e^{-4\phi} E^{aj},
\label{ar}
\een
\be
\cS_{a} = 4 e^{-2\phi} \left [
{N E_{a} \over \sqrt{h}} + 2{N e^{-2\phi} E_{aj} A^{j} \over \sqrt{h}}
+ \na_{a}(N A_{0}) + \LieN A_{a} \right ],
\label{as}
\ee
\be
\cT^{ij} = - 2 \na_{a} (N \sqrt{h} e^{-4\phi} H^{aij}) - 2 \LieN
(e^{-4\phi} E^{ij}),
\label{at}
\ee
\be
\cU_{ij} = {2 N e^{-4\phi} \over \sqrt{h}}
\left ( E_{ij} + 4 E_{i} A_{j} \right ) +
2 e^{-4\phi} \left [ 2 (\LieN A_{i}) A_{j} + \LieN B_{ij} \right ]
- 4 e^{-4\phi} \na_{i} (N B_{j0}),
\label{au}
\ee
\ben
\cY = &-& 4\na_{a}(N^{a} E) - {4 e^{-5\phi} N E_{ij} E^{ij} \over
\sqrt{h}} - 2 N \sqrt{h} e^{-3\phi}F_{ab}F^{ab} - {4 e^{-3\phi}N E_{i}
E^{i} \over \sqrt{h}} + \\ \nonumber
&-& {4 \over 3}e^{-5\phi} N \sqrt{h} H_{ijk}
H^{ijk} - 32 {e^{-5\phi} N E_{i}E^{ij}A_{j} \sqrt{h}} + \\ \nonumber
&-& 8 \left (
e^{-3\phi}E^{i} \LieN A_{i} + e^{-5\phi} E^{ij} \LieN B_{ij} +
e^{-3\phi} E^{i} \na_{i} (N A_{0}) \right ) + \\ \nonumber
&+& 16 \left (
e^{-5\phi}E^{ij} \na_{i}(N
B_{j0}) + e^{-5\phi} E^{ij} A_{j} \LieN A_{i}
\right ),
\label{ay}
\een
\be
\cZ = 4 \left ( {N E \over \sqrt{h}} + \LieN \phi \right ).
\label{az}
\ee
In Eq.(\ref{ap}) the form of $a^{ab}$ is given as follows
\ben \label{aa}
a^{ab} &=& {1 \over h} \left [
\left ( 2 \pi^{a}{}{}_{j} \pi^{bj} - \pi \pi^{ab} \right ) -
{1 \over 2}h^{ab} \left (\pi_{ij} \pi^{ij} - {1 \over 2} \pi^{2} \right
) \right ] +
\left (R^{ab} - {1 \over 2}h^{ab} R \right )
- {E^{2} h^{ab} \over h} \\ \nonumber
&+& {2 e^{-4\phi} \over h}
\left ( E^{aj} E_{j}{}{}^{b} - {1 \over 4}h^{ab} E_{ij} E^{ij} \right )
+ {2e^{-2\phi} \over h} \left (
E^{a}E^{b} - {1 \over 2}h^{ab} E_{i}E^{i} \right ) + \\ \nonumber
&+& {4e^{-4\phi} \over h} \left (
2E^{a}E^{bj}A_{j} - 2E_{i}E^{ia}A^{b} - h^{ab}E_{i}E^{ij}A_{j}
\right ) + \\ \nonumber
&+& \left (
\na_{i} \phi \na^{i} \phi h^{ab} - 2 \na^{a} \phi \na^{b} \phi
\right ) +
2e^{-2\phi} \left (
F^{aj}F_{j}{}{}^{b} + {1 \over 4}h^{ab} F_{ij}F^{ij} \right ) +
e^{-4\phi} \left (
- H^{aij}H_{ij}{}{}^{b} + {1 \over 6}h^{ab}H_{ijk}H^{ijk} \right ).
\een 
In Eqs.(\ref{ap})-(\ref{az}) the symbol $\LieN$ stands for the 
Lie derivative taken on the hypersurface $\Si$, with respect to
the vector field $N^{i}$. The Lie derivative of
$h_{ab}, A_{i}, B_{ij}$ and $\phi$ are the ordinary Lie derivatives,
while the Lie derivative of $E_{a}, E_{ij}, E^{ij}A_{j}, \pi^{ab}$
are understood as the Lie derivatives of the adequate tensor densities.
\par
The evolution equations for the EMAD system
can be obtained from Eq.(\ref{ha}), by means of the Hamiltonian's
principle, taking variations of compact support of the hypersurface
$\Si$. Then, one establishes
\be
\dot \pi^{ab} = - \cP^{ab},
\label{a141}
\ee
\be
\dot h_{ab} = \cQ_{ab},
\label{a142}
\ee
\be
\dot \pi^{(A) a} = - \cR^{a},
\label{a143}
\ee
\be
\dot A_{a} = \cS_{a},
\label{a144}
\ee
\be
\dot \pi^{(H)ij} = - \cT^{ij},
\label{a145}
\ee
\be
\dot B_{ij} = \cU_{ij},
\label{a146}
\ee
\be
\dot \pi^{(\phi)} = - \cY,
\label{a147}
\ee
\be
\dot \phi = \cZ.
\label{a148}
\ee
As was pointed by Regge and Teitelboim \cite{re,han}, 
Eq.(\ref{ha}) depicts
rather the volume contribution to the Hamiltonian and when one considers
the perturbation to the Hamiltonian $H_{V}$ satysfying asymptotic
boundary conditions at infinity, the nonvanishing surface terms arise
due to integration by parts, in order to put $\de H_{V}$ in the form of
Eq.(\ref{a17}). These surface terms can be excluded by adding the
additional surface terms \cite{re}. The result may be written as
\ben \label{a150}
H &=& H_{V} + \int_{S ^{\infty}} dS_{i}
\bigg [
N \left ( \na^{a}h_{a}{}{}^{i} - \na_{i} h_{m}{}{}^{m} \right ) +
{2 N^{b} \pi^{i}{}{}_{b} \over \sqrt{h}} + 4 e^{-4\phi}
N^{a}A_{a}E^{ij}A_{j} + \\ \nonumber
&+& 4 e^{-2\phi} \left (NA_{0} + N^{a}A_{a} \right ) E^{i} -
4 e^{-4\phi} \left (NB_{j0} + N^{a}B_{ja} \right ) E^{ij} \bigg ].
\een
A direct calculation can visualize that for all asymptotically
flat perturbations and for $N^{\mu}, A_{0}, B_{j0}$ satysfying
adequate asymptotic conditions at infinity, one arrives at
\be
\de H = \ints \left ( \cP^{ab} \de h_{ab} +
\cQ_{ab} \de \pi^{ab} + \cR^{a} \de A_{a} + \cS_{a} \de E^{a} +
\cT^{ij} \de B_{ij} + \cU_{ij} \de E^{ij} + \cY \de \phi + \cZ \de E \right ).
\label{hah}
\ee

\section{Extrema of Mass and first law of black holes mechanics}

One can define \cite{su} the canonical energy on the constraint submanifold of
the phase space as the Hamiltonian function corresponding
to the case when $N^{\mu}$ is an asymptotical translation at infinity
(i.e., $N \rightarrow 1, N^{a} \rightarrow 0$). From Eq.(\ref{a150}),
one obtains 
\be
\cE = m + \cE^{(\phi-F)} + \cE^{(\phi-B)},
\label{e}
\ee
where $m$ is the ADM mass, defined as
\be
m = {1 \over 16\pi} \int_{S^{\infty}}dS_{a} \left (
\na_{b}h^{ab} - \na^{a}h_{b}{}{}^{b} \right ),
\ee
The quantities $\cE^{(\phi-F)}$ and $\cE^{(\phi-B)}$ are equal to
\be
\cE^{(\phi-B)} =
{1 \over 4 \pi} \int_{S^{\infty}}dS_{i} e^{-4\phi}B_{0j} E^{ij},
\label{efb}
\ee
\be
\cE^{(\phi-F)} =
{1 \over 4\pi} \int_{S^{\infty}}dS_{i} e^{-2\phi}A_{0}E^{i}.
\label{eff}
\ee
In the stationary case, $A_{0}$ is uniquely determined up to a 
time-in\-depen\-dent gauge transformation, by the condition $\dot A_{a} = 0$
and $\dot \pi^{(A)i} = 0$ for all time when $N^{\mu}$ is taken to be
stationary Killing field. One can show \cite{yam}, 
that these conditions lead
to the relation
\be
\cE^{(\phi-F)} = V_{(F)} Q^{(\phi-F)},
\ee
where $V_{(F)} = (A_{0} A_{0})^{1 \over 2}$ and 
\be
Q^{(\phi-F)} =
\pm {1 \over 4\pi} \int_{S^{\infty}} e^{-2\phi} \mid E^{a} r_{a} \mid
dS,
\label{df}
\ee
$r_{a}$ is the unit radial vector in the metric $\de_{ab}$.
By analogy with the notion of the Yang-Mills charge \cite{yam}, we
will call expression (\ref{df}) as a {\it dilaton-electric} charge.
\par
In the case of the quantity $\cE^{(\phi-B)}$ the situation is similar. A
time-independent gauge transformation yields that $\dot B_{ij} = 0$,
$\dot \pi^{(H)ij} = 0$. Taking into account Eq.(\ref{au})
one can show that contracting Eq.(\ref{a145}) with $B_{j0}$, we reach to the
condition of the asymptotical constancy of the magnitude of $B_{j0}$
\be
V_{(B)} = \lim \limits_{r \rightarrow \infty} (B_{j0} B_{j0})^{1 \over
2}. 
\ee
In addition considering the condition $\dot \pi^{(H)ij} = 0$, having
in mind the asymptotical behavior of $A_{i}, B_{ij}, E_{i}, E_{ij}$,
one concludes
that
\be
\cE^{(\phi-B)} = V_{(B)} Q^{(\phi-B)},
\ee
where 
\be
Q^{(\phi-B)} =
\pm {1 \over 4\pi} \int_{S^{\infty}} e^{-4\phi} \mid E^{ij} r_{i} r_{j}
\mid dS.
\label{db}
\ee
Using the definition of $E_{ij}$, Eq.(\ref{a8}), we see that because of
the skew symmetricity of $E_{ij}$, expression (\ref{db}) is equal
to zero.

\par
We choose $N^{\mu}$ to be the stationary Killing vector field and
select $A_{0}$ and $B_{j0}$ in a manner that, $A_{a}, E_{i}, B_{ij},
E_{ij}, \phi, E$ are time independent. This choice makes the right-hand
side of Eq.(\ref{hah}) vanish. Having in mind expressions
(\ref{e}), (\ref{efb}) and (\ref{eff}) we reach to
the results
which in some aspects generalize the theorem revealed by Sudarsky and
Wald \cite{su} in the case of EYM gravity, to the case of EMAD gravity.
\par

\noindent
{\bf Theorem}\\
\noindent
Consider $(h_{ab}, \pi^{ab}, A_{a}, E^{a}, B_{ij}, E^{ij}, \phi, E)$ 
to be
smooth data for a stationary asymptotical flat solution of the
EMAD gravity. The initial data hypersurface
$\Si$ has only one asymptotic region and has no
interior boundary. Moreover, consider 
$(\de h_{ab}, \de \pi^{ab}, \de A_{a}, \de E^{a}, \de B_{ij}, 
\de E^{ij}, \de \phi, \de E)$ to be an arbitrary smooth asymptotically
flat solution of the linearized constraint equations. Then,
the following is satisfied:
\be
0 = \de \cE = \de m + V_{(F)} \de Q^{(\phi-F)}.
\label{t1}
\ee
\par
From Eq.(\ref{t1}) we see that every stationary solution to
EMAD gravity is an extremum of the ADM mass
at fixed {\it dilaton-electric} charge
described by Eq.(\ref{df}).

\vspace{0.3cm}
In the same way, one can define the canonical angular momentum $\cJ$
on the constraint submanifold of the phase space to be the Hamiltonian
function corresponding to the case where $N^{\mu}$ is an
asymptotic rotation at infinity (i.e. $N \rightarrow 0, N^{a}
\rightarrow \phi^{a}$)
\be
\cJ = - {1 \over 16 \pi} \int_{S^{\infty}}dS_{i}
\left (
2 \phi^{b} \pi^{i}{}{}_{b} + 4e^{-2\phi} \phi^{a}A_{a} E^{i} +
4e^{-4\phi} \phi^{a}B_{aj}E^{ij} + 4e^{-4\phi} \phi^{a}A_{a}E^{ij}A_{j}
\right ).
\label{js}
\ee
Converting the surface integral in Eq.(\ref{js}) and using the constraint
equation (\ref{a16}), one gets
\be
\cJ = - {1 \over 16 \pi} \ints \big ( 
4 E \Lief \phi + \pi^{ab} \Lief h_{ab} + 4 e^{-2\phi} E^{i} \Lief A_{i}
+ 2 e^{-4\phi} E^{ij} \Lief B_{ij} + 4 e^{-4\phi}E^{ij} A_{j} \Lief
A_{i} \big ).
\ee
The integral over the hypersurface $\Si$ disappear becuse of the fact
that the axial Killing field $\phi^{\mu}$ is equal to its tangential
projection. Thus, in the case when $\Si$ is a three-dimensional manifold
without boundary $\cJ = 0$, for any axisymmetric solution.
\par
Now, let us consider the case when $\Si$ has an asymptotic region and a
smooth interior boundary $S$. As in Ref.\cite{su}, we will be 
mostly interested when
$N^{\mu}$ asymptotically reaches a linear combination of a time
translation and rotation at infinity (i.e. $N \rightarrow 1,
N^{a} \rightarrow \Omega \phi^{a}$, $\phi^{a}$ is an axial Killing
field, $\Omega$ is a constant). One readily finds
\ben \label{tt}
16 \pi \big ( \de \cE - \Omega \de \cJ \big ) &=& 
\ints \left ( \cP^{ab} \de h_{ab} +
\cQ_{ab} \de \pi^{ab} + \cR^{a} \de A_{a} + \cS_{a} \de E^{a} +
\cT^{ij} \de B_{ij} + \cU_{ij} \de E^{ij} + \cY \de \phi + \cZ \de E 
\right ) \\ \nonumber
&+& \de (surface \quad terms).
\een
We choose the hypersurface $\Si$ to be an asymptotically flat one which
intersects the bifurcation sphere $S$ of the stationary black hole.
Then, we set $N^{\mu} = \chi^{\mu} = t^{\mu} + \Omega \phi^{\mu}$ and
select $A_{0}$ and $B_{j0}$ so that $\dot A_{a} = \dot E_{i} =
\dot B_{ij} = \dot E_{ij} = \dot \phi = \dot E = 0$.
By means of Eqs.(\ref{a141})-(\ref{a148}) the first 
integral on the right-hand side of Eq.(\ref{tt}) vanishes.
All but one surface terms
will also be equal to zero because of the fact that $N^{\mu} =
0$ on $S$. The nonzero term \cite{su} is given as follows:
\be
\int_{S} dS_{a} \na_{b}N \de h_{cd} (h^{ac}h^{bd} - h^{ab}h^{cd}) =
2 \kappa \de A,
\ee
where $\kappa$ is the surface gravity, constant over $S$, $A$ is the
area of $S$.
Having in mind (\ref{tt}) and (\ref{t1})
we reach to the conclusion.
\par
\noindent
{\bf Theorem}\\
Let $(h_{ab}, \pi^{ab}, A_{a}, E^{a}, B_{ij}, E^{ij}, \phi, E)$ on a
hypersurface be smooth asymptotically flat initial data for a
stationary black hole with bifurcation sphere lying on $\Si$.
Moreover, let
$(\de h_{ab}, \de \pi^{ab}, \de A_{a}, \de E^{a}, \de B_{ij}, 
\de E^{ij}, \de \phi, \de E)$ be an arbitrary smooth asymptotically
flat solution of the linearized constraint equations. Then,
the following is satisfied:
\be
\de \cE = \de m + V_{(F)} \de Q^{(\phi-F)} - \Omega
\de \cJ = {1 \over 8\pi} \kappa \de A.
\label{t2}
\ee
Equation (\ref{t2}) constitutes the extension of the first law of black
hole mechanics to the case of EMAD theory. 
For the EYM case this law was derived by Sudarsky and
Wald \cite{su}. As was stated this derivation is true for arbitrary
asymptotical flat perturbations of a stationary black hole not merely
for perturbations to other stationary black holes as was done in the
original derivation provided by Bardeen, Carter and Hawking \cite{bch}.
\par
In Ref.\cite{l1}, Gibbons {\it et al.} considered the thermodynamical
properties of black holes in the string theory. It turned out that,
these properties are conditioned on the values of certain massless
scalar fields (referred to as moduli fields) at spatial infinity. The
authors stated that, for black holes in the string theory the
dependence on the scalar charge would not vanish in the general case.
However, when
moduli fields at spatial infinity are chosen to extremize the ADM
mass at fixed entropy, angular momentum and conserved electric and
magnetic charges, this dependence vanishes. Despite the extra term in
the first law of black holes mechanics, the integrated version of it
(the Smarr formula) is lack of the dependence on the scalar charge.
\par
The EMAD gravity theory can be understand as an $SL(2,R)$ sigma model
coupled to a vector field and gravity, so one can verify our results
invoking attitude presented in Ref.\cite{l1}.
Namely, from Eq.(\ref{t2}) we can conclude that any stationary
black hole solution to the EMAD gravity, with
bifurcate Killing horizon is an extremum of the ADM mass at fixed
{\it dilaton-electric} charge, canonical
angular momentum and horizon area.
\par
In order to check this assertion one ought to consider a simple example
of a stationary black hole in the theory under consideration. We
postpone this problem to consider in a separate publication elsewhere.

\par
\vspace{0.4cm}
Now, we focuse our attention on the problem if the converse 
results to Eqs.(\ref{t1}) and (\ref{t2}) hold. Namely, whether the 
initial data for a stationary solution which are an extremum of the
ADM mass $m$ at fixed $Q^{(\phi-F)}$ are
necessary ones and if the initial data which are the extremum
of $m$ at fixed $Q^{(\phi-F)}$, angular momentum
and the horizon area are obligatory initial data for a stationary
black hole solution. In EYM gravity this problem was widely elaborated
by Sudarsky and Wald \cite{su}. We will follow this line of reasoning
in the direction of proving the converse theorems. Of course, 
one should be aware
that the argumentation is not the complete proof of the conversed
theorem and generalization to the asymptotically flat EMAD case should
be provided, giving necessary and sufficient conditions for solving
the adequate equations for perturbations on a compact support.
\par
Consider any EMAD initial data satysfying the EMAD constraint Eqs.
(\ref{a15})-(\ref{b}). Suppose, that we have smooth perturbed initial
data satisfying the linearized EMAD constraints in the
neighborhood of infinity satisfying $\tde Q^{(\phi-F)} = 0, 
\tde m \neq 0$. One has to solve the following
relations:
\be
\de C_{\mu} = S_{\mu},
\label{co1}
\ee
\be
\de C_{i} = S_{i},
\label{co2}
\ee
\be
\de C = S,
\label{co3}
\ee
where $S_{\mu} = - \tde C_{\mu}, S_{i} = - \tde C_{i}, S = - \tde C$.
Perturbations $(\de h_{ab}, \de \pi^{ab}, \de A_{a}, \de E^{a}, 
\de B_{ij}, \de E^{ij}, \de \phi, \de E)$ 
fall off sufficiently rapidly at infinity, that the relations
$\de m = \de Q^{(\phi-F)} = 0$ are satisfied.
\par
We will generalize the arguments provided by Fisher and Marsden
\cite{fis}, who studied the linear stability, in the vacuum case,
of compact ( without boundary ) Cauchy hypersurface $\Si$. They proved
that an equation of the form of Eq.(\ref{co1}) can 
be solved for a given smooth source
$S_{\mu}$ iff $S_{\mu}$ is orthogonal to the kernel of the $L^{2}$
adjoint operator $\cA^{\dagger}$, of the form $\cA(\de h_{ab}, \de
\pi^{ab}) = \de C_{\mu}$.\par
In our case,
the left-hand sides of Eqs.(\ref{co1})-(\ref{co3}) defined the
operator $\cB$ which mapped perturbations of EMAD
initial data into covariant vector, scalar fields on the hypersurface
$\Si$.  In order to find to find the adjoint operator $\cB^{\dagger}$
we multiply Eq.(\ref{co1}) by $M^{\mu}$, Eq.(\ref{co2}) by $a^{i}$
and Eq.(\ref{co3}) by $a$, and integrate over hypersurface $\Si$.
We remark that the equation
\be
\de H_{V} = \ints \bigg (
N^{\mu} \de C_{\mu} + N^{\mu} A_{\mu} \de \tiA + N^{\mu} B_{j \mu} \de
\tiB^{j} \bigg ),
\label{adj}
\ee
effectively computes the adjoint operator $\cB^{\dagger}$ and Eqs.
(\ref{a141})-(\ref{a148}) show that $M^{\mu}$ lies in the kernel of
the adjoint operator if $M^{\mu}$ is a Killing vector field for the
background initial data and
\ben \label{gau1}
\dot A_{a} = \LieM A_{a} = 0, \\ \nonumber
\dot E^{a} = \LieM E^{a} = 0, \\ \nonumber
\dot B_{ij} = \LieM B_{ij} = 0, \\ \nonumber
\dot E^{ij} = \LieM E^{ij} = 0, \\ \nonumber
\dot \phi = \LieM \phi = 0, \\ \nonumber
\dot E = \LieM E = 0 ,
\een
in a guage where the following is satisfied
\ben \label{gau2}
M^{\mu} A_{\mu} = a, \\ \nonumber
M^{\mu} B_{j \mu} = a_{j}.
\een
By the analogy with the vacuum case, we conjecture that Eqs.
(\ref{co1})-(\ref{co3}) can be solved if there do not exist
$(M^{\mu}, a^{j}, a)$  satysfying asymptotical conditions, where
$M^{\mu}$ is a Killing vector field for the background spacetime
and Eqs.(\ref{gau1}) and (\ref{gau2}) are satisfied. Then, a necessary
condition for an extremum of mass at fixed charges is that the
background EMAD initial data correspond to a stationary solution.
\par
On the other hand, when we take into consideration the case with the
boundary $S$, one should begin with the perturbations
$(\tde h_{ab}, \tde \pi^{ab}, \tde A_{a}, \tde E^{a}, 
\tde B_{ij}, \tde E^{ij}, \tde \phi, \tde E)$ satisfying the
constraints near infinity and also satisfying
$\tde Q^{(\phi-F)} = \tde \cJ = \tde A = 0$ and
$\tde m \neq 0$. Once more, we should solve Eqs.
(\ref{co1})-(\ref{co3}) with the additional conditions that
$\de Q^{(\phi-F)} = \de \cJ = \de A = 0$.
As in the previous case the background spacetime must admit a Killing
field $M^{\mu}$ and $a^{j}, a$ satisfying Eqs.(\ref{gau1})-(\ref{gau2}).
In order to have no surface terms, the requirement revealed from the
condition  $\de \cJ = 0$, is that $M^{a} \rightarrow \phi^{a}$,
$M^{0} \rightarrow const$, $a^{j} \rightarrow const$ and
$a \rightarrow const$ at infinity. The requirement that
$M^{\mu} = 0$ at $S$ and $\na_{a} M^{0}$ has a constant magnitude on
$S$ assures us that no surface terms will be generated at $S$.
As was pointed in Ref.\cite{su}, the fact that the Killing vector field
$M^{\mu}$ vanishes at $S$ caused that the null geodesics orthogonal to
the surface $S$ generate a bifurcate horizon.
\par
The nonexistence of the triple $(M^{\mu}, a^{j}, a)$ fulfilling the
discussed asymptotic conditions at infinity and boundary conditions at
the surface $S$, is obligatory to solve Eqs. (\ref{co1})-(\ref{co3})
with arbitrary smooth source terms of compact support for
$\de \cJ = \de Q^{(\phi-F)} = \de m = \de A = 0$.
Then, the converse theorem to the theorem (\ref{t2}) takes place;
namely in order for EMAD initial data to be an extremum of mass at
fixed {\it dilaton-electric} charge,
canonical momentum and horizon area, the initial data 
have to coincide with a
stationary black hole with a bifurcation surface and a bifurcate 
Killing horizon.

\section{The mass formulas and the staticity problem for nonrotating
black holes in Einstein-Maxwell axion-dilaton gravity}

Following the attitude presented by Sudarsky and Wald 
\cite{su1},
we will deal in this section with the problem of 
finding the conditions for nonrotating 
EMAD black holes to be static.
\par
In EYM gravity Sudarsky and Wald \cite{su} were able to give the 
proof of staticity theorem for nonrotating black holes which 
with no additional inequalities as was achieved in the
first proof provided by Carter \cite{ca0,ca}. 
Now, we will try to find the simillar 
way of getting rid of the additional assumption in the staticity theorem.
\par
We will consider the spacetime of 
a stationary black hole with bifurcate Killing horizon.
This kind of black holes possesses a Killing vector field $t^{\mu}$,
which becomes a time translation in the asymptotic region and a Killing
vector field $\chi^{\mu}$. The vector field $\chi^{\mu}$ vanishes
on the bifurcation sphere. If $\chi^{\mu}$ does not coincide with the 
$t^{\mu}$ field, then the spacetime has an axial Killing field,
$\phi^{\mu}$ fulfilling the relation $ \chi^{\mu} = t^{\mu} +
\Omega \phi^{\mu}$, where $\Omega$ is constant known as the angular
velocity.
\par
Chru\'sciel and Wald \cite{ch} proved that any 
stationary black hole with bifurcate
Killing horizon admits an asymptotically flat maximal hypersurface
which is asymptotically orthogonal to $t^{\mu}$ and its boundary is the 
bifurcation surface $S$, of the horizon. 
In what follows, one will take into account the 
hypersurface $\Si$ to be such a maximal hypersurface.
\par
Considering the initial data which are induced on $\Si$ and 
choosing the lapse and shift function to coincide with
a Killing field in the spacetime, the Eqs.(\ref{a141})-(\ref{a148}) have
their right-hand sides equal to zero. Then, contracting Eq.(\ref{a141})
we reach to the expression
\be
\na_{i} \na^{i} N = \rho N,
\label{d1}
\ee
where $\rho$ is the non-negative quantity in the form as follows
\be
\rho = {1 \over h}\pi_{ab} \pi^{ab} +
{2 E^{2} \over h} + {e^{-2 \phi} E_{i}E^{i} \over h} +
{8e^{-4\phi} E_{i}E^{ij}A_{j} \over h} + {2 e^{-4\phi}E_{ij}E^{ij}
\over h}
+ {1 \over 2}e^{-2\phi}F_{ab}F^{ab} + {1 \over 3} e^{-4\phi}H_{ijk}H^{ijk},
\label{d2}
\ee
Next we use
the lapse function in the form defined by $\lambda = -
n_{\alpha}t^{\alpha}$, where $t^{\mu}$ is the stationary Killing vector
field.
The boundary conditions for the lapse function are
$\lambda \vert_{S} = 0, \lambda \vert_{\infty} = 1$.
Integrating Eq.(\ref{d1}) over $\Si$, taking into account that
the surface integral over $S^{\infty}$ is $4\pi M$ and the surface
integral over $S$ is equal to $\kappa A$, one can reach to 
the mass formula in EMAD gravity, namely,
\be
4 \pi M - \kappa A = \ints \lambda \rho.
\label{d3}
\ee
To proceed further, we take into account mass formula obtained by Bardeen,
Carter and Hawking \cite{bch}
\be
M - \kappa A - 2 \Omega J_{H} = 2 \ints \left (
T_{\mu \nu} - {1 \over 2}T g_{\mu \nu} \right ) t^{\mu} n^{\nu},
\label{d4}
\ee
where $J_{H}$ is the angular momentum of the black hole, defined \cite{wa}
in the standard way by $J_{H} = {1 \over 16 \pi} \int_{S} 
\ep_{\alpha \beta \ga
\de} \na^{\ga} \phi^{\de} $.
\par
In EMAD the explicit form of Eq.(\ref{d4}) is
\ben \label{d5}
M - \kappa A - 2 \Omega J_{H} &=& 2 \ints \bigg [
{4 \la E^2 \over h} + 4 t^{m}\na_{m} \phi {E \over \sqrt{h}} +
e^{-2\phi} \left (
{2 \la E_{a}E^{a} \over h} + 4t^{m} F_{md} {E^{d} \over \sqrt{h}} +
\la F_{ab}F^{ab} \right ) +
\\ \nonumber
&+& 2e^{-4\phi} \left (
t^{m}H_{mij} {E^{ij} \over \sqrt{h}} + {2 \over 3}{\la E_{ij}E^{ij}
\over h} + {1 \over 3} \la H_{ijk}H^{ijk}
\right ) \bigg ]
\een
As in Ref.\cite{su1},
one can show that using the definition of $\cJn$
and changing the surface integral into a volume integral over the
hypersurface $\Si$, taking into account the constraint, Eq.
(\ref{a16}), and the fact that the integral over $\Si$ in the definition
of $\cJn$ vanishes because of the fact that the axial Killing
field $\phi^{\mu}$ is equal to its tangential projection $\phi^{i}$,
one gets that
\be
\cJn = \cJ_{H}.
\label{d6}
\ee
Recalling that the first term in equation for $\cJ_{H}$ is equal
to $J_{H}$ \cite{su1} and computing the other terms, having in mind that
on $S$ vectors $t^{a}$ and $\phi^{a}$ coincide up to the 
constant $\Omega$ as the result of vanishing $\chi^{a}$ on $S$,
one readily finds
\be
4 \pi ( J_{H} - \cJn ) \Omega =
- \int_{S} {dS_{i} \over \sqrt{h}} \bigg (
t^{m} A_{m}e^{-2\phi} E^{i} + t^{m}B_{mj} e^{-4\phi} E^{ij} +
t^{m}A_{m}e^{-4\phi}E^{ij}A_{j} \bigg ).
\label{d7}
\ee
Now, we change the surface integral into the volume one and we
take into consideration the asymptotic behavior of $A_{i}, E^{i},
B_{ij}, E^{ij}$ at infinity. One can draw a conclusion that there will
be no contribution from the boundary at infinity. Thus,
from equation (\ref{d7}) we arrive at the following relation:
\ben
4 \pi ( J_{H} - \cJn ) \Omega &=& \ints \bigg [
t^{m}F_{im}e^{-2\phi} E^{i} + e^{-2\phi}E^{a} \Liet A_{a} +
e^{-4\phi} \left (
{1 \over 2} E^{ab} \Liet B_{ab} + E^{cd} A_{d} \Liet A_{c} \right )
+ \\ \nonumber
&-& {1 \over 2}t^{m}e^{-4\phi} H_{mij} E^{ij} \bigg ].
\label{d8}
\een
We use Eqs.(\ref{a144}) 
and (\ref{a146}),
$\dot B_{ij} = 0$ and $\dot A_{i} = 0$,
for the explicit form of $\LieN B_{ij}$ and
$\LieN A_{b}$. Of course, there will be no contribution from the
surface integrals over $S$, because of the boundary conditions for the
lapse function $\la$. Then, we get
\ben \label{d9}
4 \pi ( J_{H} - \cJn ) \Omega &=& \ints \bigg [
- {\la e^{-2\phi}E_{a}E^{a} \over \sqrt{h}} - 4{\la
e^{-4\phi}E_{i}E^{ij} A_{j} \over \sqrt{h}} - {\la
e^{-4\phi}E_{ab}E^{ab} \over \sqrt{h}} + \\ \nonumber
&+& t^{m} \bigg (
F_{im}e^{-2\phi}E^{i} - {1 \over 2}e^{-4\phi}H_{mij}E^{ij}
\bigg ) \bigg ] + \\ \nonumber
&-& \int_{S^{\infty}} dS_{i} \ \la A_{0} e^{-2\phi} E^{i} -
\int_{S^{\infty}} dS_{i} \ \la B_{0j} e^{-4\phi} E^{ij}.
\een
Eliminating, from relation (\ref{d9}), $J_{H}$ by means of Eq.(\ref{d5})
\ben \label{d10}
4 \pi M &-& \kappa A + 8 \pi
V_{(F)} Q^{(\phi-F)}
+ 8\pi \Omega \cJn = \\\ \nonumber
&=& \ints \ \la \bigg (
{1 \over 2} e^{-2\phi}F_{ab}F^{ab} + {1 \over 3}e^{-4\phi}H_{ijk}
H^{ijk} - {e^{-2\phi} E_{a}E^{a} \over h}
- 8 {e^{-4\phi}E_{i}E^{ij}A_{j} \over h} - 2 {e^{-4\phi} E_{ij}E^{ij}
\over h} \bigg ).
\een
Taking into considerations Eqs.(\ref{d10}), (\ref{d2}) and
(\ref{d3}), one can conclude
\ben \label{d11}
8 \pi \bigg ( \Omega \cJn &-& V_{(F)} Q^{(\phi-F)}
\bigg ) = \\ \nonumber
&=& \ints \ \la \bigg (
{\pi_{ab} \pi^{ab} \over h} + 2 {e^{-2\phi} E_{a}E^{a} \over h} +
4 {e^{-4\phi}E_{ij} E^{ij} \over h} + 16 {e^{-4\phi} E_{i}E^{ij}A_{j}
\over h} \bigg ).
\een
Theorem 4.2 in Ref.\cite{ch} states 
that the exterior region of the black hole can be
foliated by maximal hypersurfaces with boundary $S$ which are
asymptotically orthogonal to the timelike Killing vector field
$t^{\mu}$, if the strong energy condition $R_{\mu \nu}Z^{\mu}Z^{\nu}
\ge 0$ for all timelike vectors $Z^{\mu}$ is fulfilled. By the direct
calculation, one can check that this is the case in
EMAD gravity. Taking the above, one can
reach to the main result of this section, namely, the following.
\par
\noindent
{\bf Theorem} \\
Consider an asymptotically flat solution to
EMAD gravity which has a Killing vector field
which is timelike at infinity, describing a stationary black hole
comprising a bifurcate Killing horizon with a bifurcation surface $S$.
Suppose, moreover, that the following is satisfied:
$$\Omega \cJn - V_{(F)} Q^{(\phi-F)} = 0.$$
Then, the solution is static and has vanishing $E_{i}$ and $E_{ij}$
on the static hypersurfaces.
\par
We remark that, in the above formulation of the staticity theorem for
non-rotating black holes in EMAD gravity, we managed to get rid of the
additional assumption used in the previous proof of the staticity
theorem in EM gravity \cite{ca0,ca} and in the prior attitude to the
staticity problem in EMAD gravity \cite{rog3}.





\end{document}